# Hamiltonian Reordering for Shallower Trotterization Circuits

**Cédric Ho Thanh**

cedric.hothanh@riken.jp

Advanced Data Science Project, RIKEN Information R&D and Strategy Headquarters, RIKEN, Tokyo, Japan

March 14, 2025

**ABSTRACT**

Quantum simulation is a popular application of quantum computing, but its practical realization is hindered by the technical limitations of current devices. In this work, we focus on preprocessing Hamiltonians before Trotterization to generate shallower evolution circuits, which are less prone to noise and decoherence. Specifically, we apply graph coloring techniques to reorder Pauli terms and increase "gate parallelism". We benchmark two coloring algorithms, and report the depth reduction and computational overhead. Then, we examine how these optimized circuits affect the performance of the Quantum Approximate Optimization Algorithm (QAOA). Our results show that shallower circuits lead to faster convergence and reach higher energy levels compared to their non-reordered counterparts.

## 1 Introduction

### 1.1 Background

Quantum computing has emerged as a powerful framework for solving certain classes of computational problems that would otherwise be impractical for classical computers. At the forefront of the techniques it enables is *quantum simulation*. If the laws of a system in initial state $|\psi(0)\rangle$ are formulated in terms of a time-independent Hamiltonian $H$, then the Schrödinger equation dictates that the state of the system at time $t \geq 0$ is given by $|\psi(t)\rangle = e^{-itH}|\psi(0)\rangle$. Provided with a quantum circuit that faithfully implements the evolution operator $e^{-itH}$, it is thus possible to simulate that system at any time $t$. This is particularly relevant in condensed matter physics, quantum chemistry, and discrete optimization problems [1].

However, implementing $e^{-itH}$ from $H$ is a non-trivial task. A ubiquitous technique is called *Trotterization* [2, 3], where given a decomposition into simple terms $H = \sum_j H_j$, the operator $e^{-itH}$ can be approximated using $e^{-i\delta t H_j}$, representing a smaller evolution of a simpler system, modulo an error which in practice is usually quadratic in the timestep $\delta t$. Usually, the $e^{-i\delta t H_j}$ are realizable in hardware, and so may be assembled into actually executable quantum circuits. However, these Trotterization circuits have a depth that increases polynomially with the complexity of $H$ and the number of time steps $N = t/\delta t$. This poses significant challenges for *Noisy Intermediate-Scale Quantum era* (NISQ) devices to execute faithfully, primarily due to hardware noise and limited coherence times.

Consequently, producing amenable Trotterization circuits is of crucial practical importance.

## 1.2 Contribution

This work is concerned with *Pauli Hamiltonians* (thereafter simply "Hamiltonians"), which can be expressed as sums of tensor products of *Pauli matrices*. These are the most common types of Hamiltonians encountered in applications of quantum computing.

In Section 3, we present a way to systematically map Hamiltonians to graph coloring problems, and how colorings can be used to restructure them before Trotterization. In addition to the general case, we also consider the special case of *Ising Hamiltonians*, widely used to model certain classes of binary optimization problems [1, 4]. Their 2-local and "non-redundant" structure allows for a more compact graph representation and transform a vertex coloring problem into an edge coloring problem.

In Section 3.5, we benchmark our approach on a subset of the HamLib dataset [5], a large repository of synthetic and real-world Hamiltonians from various fields, including binary optimization. We empirically demonstrate that our reordering approach significantly reduces Trotterization circuit depth, but at the cost of a noticeable computational overhead.

Then, in Section 4.2 we assess the impact of depth reduction in the context of the *Quantum Approximate Optimization Algorithm* (QAOA). We observe that circuits with reduced depth lead to faster convergence and produces higher energy states compared to their non-reordered baseline counterparts.

The data summarized in the present paper is available in Zenodo [6], and the code in [7] and on GitHub `https://github.com/altaris/pauli-coloring-benchmark`.

## 1.3 Related works

The use and impact of Hamiltonian reordering has been extensively studied in the literature, although, to our knowledge, not with the goal of depth minimization in mind.

A particularly prevalent idea is to regroup mutually commuting Pauli terms before Trotterization to either increase the odds of gate cancellation [8] or reduce readout error [9–16]. Indeed, Trotterization of operators obtained from mutually commuting Pauli terms carries no error (see Equation 12) and produces simpler circuits. Noticeably, [10] also uses graph theoretical methods, namely minimal clique covers.

Other reordering schemes such as *magnitude ordering* and *lexicographical ordering* also appear in [17, 18]. Both stem from the observation that the error induced by the Trotter approximation increases as the system passes through the various evolution operators in the Trotterized circuit. By either executing high-magnitude terms first, or my regrouping similar and thus highly overlapping terms together, the authors experimentally demonstrate error reduction in the simulation of fermionic systems.

Amendments to the Trotterization process itself have been proposed. For example, in [19, 20], the authors use adaptative timesteps to minimize the approximation error and develop the ADA-Trotter and tADA-Trotter algorithms.

## 1.4 Acknowledgements

This work was supported by the RIKEN TRIP initiative (RIKEN Quantum), the UTokyo Quantum Initiative, as well as the RIKEN Pioneering Project "Prediction for Science".

# 2 Preliminaries

## 2.1 Trotterization

If a quantum system is governed by a time-independent Hamiltonian $H$ and starts out in state $|\psi(0)\rangle$, then the Schrödinger equation dictates that its state at time $t > 0$ is

$$|\psi(t)\rangle = e^{-itH}|\psi(0)\rangle \quad (1)$$

modulo a global phase. In quantum computing, *Trotterization* is the process of approximating the evolution operator $e^{-itH}$ from $H$ by a sequence of *gates* that can be implemented on hardware. If $H$ can be decomposed as a sum $H = \sum_j H_j$ of simple operators, then this approximation usually involves products of operators of the form $e^{-i\delta t H_j}$, for some timestep $\delta t < t$. For example, the first-order *Suzuki-Trotter expansion* [3] with $k$ timesteps reads

$$e^{-itH} \approx \left(e^{-i\frac{t}{k}H_1} e^{-i\frac{t}{k}H_2} \cdots e^{-i\frac{t}{k}H_n}\right)^k. \quad (2)$$

However, if the $H_j$'s mutually commute, then Equation 2 is actually exact. This motivated many past works to look for groups of mutually commuting operator in the decomposition of $H$ [9–16].

## 2.2 Overlapping

Hamiltonians we consider in this paper are represented as sums of tensor products of the four *Pauli matrices*:

$$I = \begin{pmatrix} 1 & \\ & 1 \end{pmatrix}, \quad X = \begin{pmatrix} & 1 \\ 1 & \end{pmatrix}, \quad Y = \begin{pmatrix} & -i \\ i & \end{pmatrix}, \quad Z = \begin{pmatrix} 1 & \\ & -1 \end{pmatrix}. \quad (3)$$

Tensor products of Pauli matrices are also known as *Pauli terms* and can be written down as *Pauli strings*. For example, in a 6-qubit system, the term on the left can be abbreviated as in the middle by omitting the $\otimes$ symbols, and in the sparse case (most components are $I$) written down more succinctly still as on the right:

$$I \otimes I \otimes X \otimes I \otimes Y \otimes Z = IIXIYZ = X_3 Y_5 Z_6. \quad (4)$$

We write $A[j]$ for the $j$-th term in a Pauli string $A$.

We say that two Pauli strings $A \neq B$ *overlap* if they there is an index $j$ such that $A[j]$ and $B[j]$ are both non-identity Pauli matrices. In other words, $A$ and $B$ act non-trivially on at least one "shared" qubit. For example $IXX$ and $YIX$ overlap, whereas $IZZ$ and $ZII$ do not.

If $A$ and $B$ don't overlap, then they commute[1] and therefore Suzuki's approximation of Equation 2 is exact:

$$e^{-it(A+B)} = e^{-itA}e^{-itB}. \quad (5)$$

In addition, non-overlapping terms can be executed in parallel in the following sense. Let $N$ be the number of qubits in the current system. Up to basis permutation, we may assume that there exists an index $1 \leq p \leq N$ such that $A[j] = I$ for all $j > p$ and $B[j] = I$ for all $j \leq p$. In particular, $A$ and $B$ decompose as

$$A = \tilde{A} \otimes I^{\otimes(N-p)}, \quad B = I^{\otimes p} \otimes \tilde{B}, \tag{6}$$

for some $p \times p$ matrix $\tilde{A}$ and $(N-p) \times (N-p)$ matrix $\tilde{B}$. Then,

$$\begin{aligned} e^{-it(A+B)} &= e^{-itA} e^{-itB} \\ &= \left( e^{-it\tilde{A}} \otimes I^{\otimes(N-p)} \right) \left( I^{\otimes p} \otimes e^{-it\tilde{B}} \right) \\ &= e^{-it\tilde{A}} \otimes e^{-it\tilde{B}}, \end{aligned} \tag{7}$$

which transforms the "two step" operator of Equation 5 into a "single step" operator. In terms of quantum circuits, this translates to a reduction in circuit depth (see Figure 1), which is the main goal of this paper.

---

[1] This is not a necessary condition however. For example, $IIX$ and $IXX$ overlap but still commute.

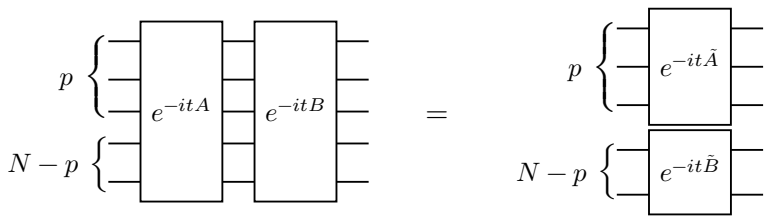

Figure 1: Gate parallelization in action

## 3 Hamiltonian Reordering

In this section, we develop Hamiltonian reordering methods and benchmark their performance on the HamLib dataset.

### 3.1 From Hamiltonians to graphs

Given a Hamiltonian $H = \sum_{j=1}^{M} H_j$ where each $H_i$ is a Pauli string, we can construct its *overlap graph* $G_{\text{overlap}} = (V, E)$ where $V = \{1, ..., M\}$ and where for $j \neq k$, there is an edge $(j, k) \in E$ if $H_j$ and $H_k$ overlap. The overlap graph is non-directed, and so the tuples $(j, k)$ and $(k, j)$ correspond to the same edge.

The overlap graph can be defined for any Hamiltonian that is a sum of Pauli string. However, in the context of *quadratic unconstrained binary optimization* (QUBO), Hamiltonians most often fall in the category of *Ising Hamiltonians* [4], which have the following form

$$H = - \sum_{1 \leq j < k \leq N} J_{j,k} Z_j Z_k - \sum_{j=1}^{N} h_j Z_j, \tag{8}$$

where $J_{j,k}, h_j \in \{0, 1\}$ and $N$ is the number of qubits. In this case, we define a finer *interaction graph* $G_{\text{inter.}} = (V, E)$, where $V = \{1, ..., N\}$, where if $J_{j,k} = 1$, then there is an edge $(j, k)$, and where if $h_j = 1$, then there is a self-loop $(j, j)$. Unlike the overlap graph, the interaction graph completely captures the structure of the Hamiltonian, in that $H$ can be reconstructed from $G_{\text{inter.}}$ (up to basis permutation). Furthermore, the number of terms in an Ising Hamiltonian $H$ can grow quadratically in the number of qubits, and so $G_{\text{inter.}}$ may offer a quadratic vertex number reduction over $G_{\text{overlap}}$.

## 3.2 Coloring of $G_{\text{overlap}}$

Consider a Hamiltonian $H = \sum_j H_j$ and its overlap graph $G_{\text{overlap}}$. By construction, two terms $H_j$ and $H_k$ overlap if and only if there is an edge between vertices $j$ and $k$. Therefore, given a vertex coloring of $G_{\text{overlap}}$, terms corresponding to vertices of the same color can be executed in parallel in the sense of Equation 7. This presumably leads to a reduction in the depth of the Trotterized circuit.

In this work, we consider the *saturation coloring* algorithm, given in Algorithm 1, which greedily assigns colors to vertices by processing them in a specific order, from most to least *saturated*. For a partially colored graph $G = (V, E)$, the *saturation* of a vertex $v$ is the number of colored neighbor of $v$. For notation convenience, we consider colors to be non-negative integers.

1:  **function** SATURATION-COLORING($G$)
2:      **while** $\exists$ uncolored vertex **do**
3:         $v \leftarrow$ uncolored vertex with max saturation
4:         color($v$) $\leftarrow$ min color not among neigh. of $v$

Algorithm 1: Greedy graph coloring with saturation heuristic

However, both $G_{\text{overlap}}$ and Algorithm 1 can be implemented more efficiently given the inherent presentation of an overlap graph. Note that for a qubit index $q$, the subgraph $K^{(q)} \subseteq G$ of all terms that act on $q$ is a complete graph (or *clique*) since all such terms mutually overlap. Furthermore, $G = \bigcup_q K^{(q)}$. Therefore, instead of representing $G_{\text{overlap}}$ as a set of vertices and edges, we may instead represent it as a sequence of (non-necessarily disjoint) sets $V^{(1)}, ..., V^{(N)}$ where $V^{(q)}$ contains all the terms acting non-trivially on qubit $q$. This allows for a more time and memory efficient reformulation of Algorithm 1 detailed in Algorithm 2

1: **function** SATURATION-COLORING($G$ presented as $(V^{(1)}, ..., V^{(N)})$)
2:     **while** $\exists$ uncolored vertex **do**
3:         ▷ *Loop invariants:*
           *1. the current (partial) coloring is valid;*
           *2. every connected component of G is either fully colored or fully uncolored*
           *3. the number of uncolored connected component is strictly decreasing*
4:         $v \leftarrow$ an uncolored vertex
5:         $\text{color}(v) \leftarrow 0$
6:         ▷ *The* fringe *$F$ is a dict. that maps vertices to sets of colors*
7:         $F \leftarrow$ empty dict.
8:         **for** $q$ such that $v \in V^{(q)}$ **do**
9:             **for** $w \in V^{(q)} - \{v\}$ **do**
10:                 $F[w] \leftarrow \{0\}$
11:         **while** $F$ is not empty **do**
12:             ▷ *Loop invariants:*
                *1. F contains all uncolored vertices adjacent to at least one colored vertex;*
                *2. if $w \in F$, then $F[w]$ is set of neighboring colors of $w$*
                *3. the current (partial) coloring is valid;*
                *4. the number of uncolored vertices is strictly decreasing*
13:             $v \leftarrow \arg\max_{w \in F} |F[w]|$
14:             $\text{color}(v) \leftarrow \min(\mathbb{N} - F[v])$
15:             **delete** $F[v]$
16:             **for** $q$ such that $v \in V^{(q)}$ **do**
17:                 **for** $w \in V^{(q)} - \{v\}$, $w$ uncolored **do**
18:                     **if** $w \in F$ **then**
19:                         $F[w] \leftarrow F[w] \cup \{\text{color}(v)\}$
20:                     **else**
21:                         $F[w] \leftarrow \{\text{color}(v)\}$

Algorithm 2: Saturation coloring algorithm with efficient clique representation of the overlap graph $G_{\text{overlap}}$

### 3.3 Coloring of $G_{\text{inter.}}$

We now consider the case of Ising Hamiltonians. Whereas terms correspond to the vertices of $G_{\text{overlap}}$, in $G_{\text{inter.}}$ they correspond to edges. By construction, two terms overlap if and only if the corresponding edges are incident, i.e. terminate at a common vertex. Therefore, given an edge coloring of $G_{\text{inter.}}$, terms corresponding to edges of the same color can be executed in parallel in the sense of Equation 7.

In this case, we consider the Misra-Gries edge coloring algorithm [21]. Its formulation in Algorithm 3 depends on a few core notions and subroutines. Let's assume that $G_{\text{inter.}}$ is uncolored or only partially colored.

A color $c$ is *free* for an edge $e = (v, w)$ if no edge incident to either $v$ or $w$ has this color. A *fan* of size $k \geq 1$ around a vertex $v$ is a sequence of edges $(e_1; e_2, ..., e_k)$ with $e_j = (v, w_j)$ for $1 \leq j \leq k$ such that: 1. all the $w_j$'s are distinct[2]; 2. $e_1$ is uncolored; 3. $e_2, ..., e_k$ are colored; and 4. the color of $e_{j+1}$ is free for $e_j$, for $1 \leq j < k$. In this context, *rotating the fan* $(e_1; e_2, ...e_k)$ consists of simultaneously assigning $\text{color}(e_j) \leftarrow \text{color}(e_{j+1})$ for $1 \leq j < k$, and un-coloring $e_k$. Note that the new coloring is still valid thanks to condition 4. In addition, the number of colored edges did not change.

---
[2]Since $G_{\text{inter}}$ is a simple graph, i.e. does not have parallel edges, this condition reduces to all the $e_j$'s being distinct.

Let $c$ and $d$ be two colors. A *c/d-path* is a maximal path of colored edges, all of which have color $c$ or $d$. If $v \in V$ is a vertex, then the *c/d/v-path* is simply a *c/d*-path that passes through $v$. For a fixed $v$, a *c/d/v-path* either does not exist (no edge incident to $v$ has color $c$ or $d$) or is unique. Given a *c/d*-path $e_1, ..., e_k$, the action of *inverting* it consists of simultaneously assigning color $d$ to all edges in the path colored with $c$ and conversely.

*Proposition*: Inverting a *c/d*-path in a valid (partial) coloring produces another valid (partial) coloring.

*Proof*: To re-evaluate the validity of the coloring, it is sufficient to check that edges incident to $v$ have different colors for all vertex $v$ in the path.

Assume that $v$ is an endpoint of the path (provided that the path is not a cycle). Say that $v$ is incident to $e_1$ and that the color of $e_1$ before inversion was $c$ (the other cases are treated similarly). By maximality, $v$ has no incident edge with color $d$. Therefore, recoloring $e_1$ with color $d$ doesn't clash with other edges incident to $v$.

Now assume that $v$ is in the interior of the path, say between $e_j$ and $e_{j+1}$ for some $1 \leq j < k$. Then since the coloring was valid before inversion, no other edge incident to $v$ had color $c$ or $d$. This still holds after inversion. □

1: **function** MISRA-GRIES-COLORING($G$)
2:     $U \leftarrow$ edges of $G$ except self-loops
3:     ▷ *Self-loops are colored at the end*
4:     **while** $U \neq \emptyset$ **do**
5:         ▷ *Loop invariants:*
            *1. the current (partial) coloring is valid;*
            *2. edges in U are uncolored*
            *3. |U| is strictly decreasing*
6:         $e_1 = (v, w_1) \leftarrow$ an edge in $U$
7:         $(e_1; e_2, ..., e_k) \leftarrow$ a max. fan around $v$ that starts with $e_1$
8:         ▷ *Write $e_j = (v, w_j)$*
9:         $c \leftarrow$ free color for $v$
10:        $d \leftarrow$ free color for $w_1$
11:        $p \leftarrow$ the unique *c/d/v*-path
12:        **if** $p$ is not empty **then**
13:            INVERT-*c/d/v*-PATH($p$)
14:        ▷ *After inversion, the fan may not be valid anymore. Specifically condition 4 may be violated.*
15:        $k \leftarrow \max\{l \mid (e_1; e_2, ..., e_l) \text{ valid fan}, d \text{ free on } w_l\}$
16:        ▷ *Such an l must exist since $(e_1;)$ is always a valid fan. In particular, $k \geq 1$.*
17:        **if** $k > 1$ **then**
18:            ROTATE-FAN($e_1; e_2, ..., e_k$)
19:        color($e_k$) $\leftarrow d$
20:        $U \leftarrow U - \{e_1\}$
21:    ▷ *Assign color to self-loops*
22:    **for** self loops $e = (v, v) \in E$ **do**
23:        ▷ *Loop invariant: the current (partial) coloring is valid*
24:        color($e$) $\leftarrow$ min color not among edges incident to $v$

Algorithm 3: Misra-Gries edge coloring algorithm, amended from [21] to handle self-loops

The Misra-Gries algorithm has a time complexity of $O(|V||E|)$, see [21], and uses at most one extra color compared to the optimal coloring.

## 3.4 Reordered Hamiltonian

Consider a Hamiltonian $H = \sum_j H_j$ and a coloring of its overlap graph $G_{\text{overlap}}$. By construction, terms (corresponding to vertices) of the same color can be parallelized in the sense of Equation 7. Therefore, we can rearrange the terms of $H$ so that its Trotterized evolution circuit is shallower. Explicitely,

$$H = \sum_j H_j = \sum_c \sum_{\text{color}(v_j)=c} H_j, \tag{9}$$

where $c$ ranges over all the colors used in the coloring of the overlap graph. Likewise, in the case of Ising Hamiltonians, an edge coloring gives a reordering

$$\begin{aligned} H &= -\sum_{j<k} J_{j,k} Z_j Z_k - \sum_j h_j Z_j \\ &= -\sum_c \sum_{\text{color}((j,k))=c} R_{j,k} Z_{j,k}, \end{aligned} \tag{10}$$

where if $j < k$, $R_{j,k} = J_{j,k}$ and $Z_{j,k} = Z_j Z_k$, and if $j = k$, $R_{j,k} = h_j$ and $Z_{j,k} = Z_j$.

## 3.5 Experiments

We test our reordering methods of Section 3.2 and Section 3.3 on the Hamiltonians of the HamLib dataset [5] representing *max-cut problems*. Given a graph $G = (V, E)$, a maximum cut is a partition $V = V_0 + V_1$ such that the number of edges between $V_0$ and $V_1$ is maximized. This is a classical NP-complete problem and one of Karp's original list [22]. Given $G$ with $V = \{1, 2, ...\}$, its max-cut problem instance can be described in therms of the following Ising Hamiltonian:[3]

$$H_G = -\sum_{(j,k) \in E} Z_j Z_k. \tag{11}$$

Given a sufficiently high energy state $|\psi\rangle$ of $H_G$, sampling a bit string $x_1...x_{|V|}$ from $|\psi\rangle$ and placing vertex $j$ in $V_{x_j}$ will produce a maximal cut of the graph with high probability.

For this benchmark, and with the goal of executing the resulting evolution circuits on a real quantum device in mind, we focus on problem instances where $32 \leq |V| \leq 127$ and $16 \leq |E| \leq 256$. We use the *4-th order Suzuki-Trotter formula with a single timestep* [3] to map max-cut Hamiltonians to evolution circuits: for $H = \sum_{j=1}^n H_j$

$$e^{-itH} \approx S_2(s_2 t)^2 \, S_2((1-4s_2)t) \, S_2(s_2 t)^2, \tag{12}$$

with

$$S_2(x) = \left(\prod_{j=1}^n e^{-i\frac{x}{2} H_j}\right) \left(\prod_{j=n}^1 e^{-i\frac{x}{2} H_j}\right) \tag{13}$$

and $s_2 = 1/\left(4 - \sqrt[3]{4}\right)$. Baseline (i.e. non-reordered) Hamiltonians produce evolution circuits with depths ranging from 45 to 2365 gates, with an average of 215.4 gates. The depth distribution is plotted in Figure 2.

We find that saturation coloring of $G_{\text{overlap}}$ produces evolution circuits that are 52.2% the depth of the baseline circuit on average, which amounts to an average depth reduction of 47.8%. For the Misra-Gries algorithm of $G_{\text{inter.}}$, the average depth reduction is 44%. These significant savings are at the cost of processing time (reordering and Trotterization), with an average of 157.2% and 186.3% relative to the baseline (Trotterization only), respectively. The depth distribution of the reordered circuits are plotted in Figure 3. In Figure 4 and Figure 5, we compare the distribution of the number of terms of the Hamiltonians in the dataset against depth and the processing time relative to the baseline.

---

[3]Note that the interaction graph of $H_G$ is in fact $G$ itself. This is not relevant for the rest of this paper, however.

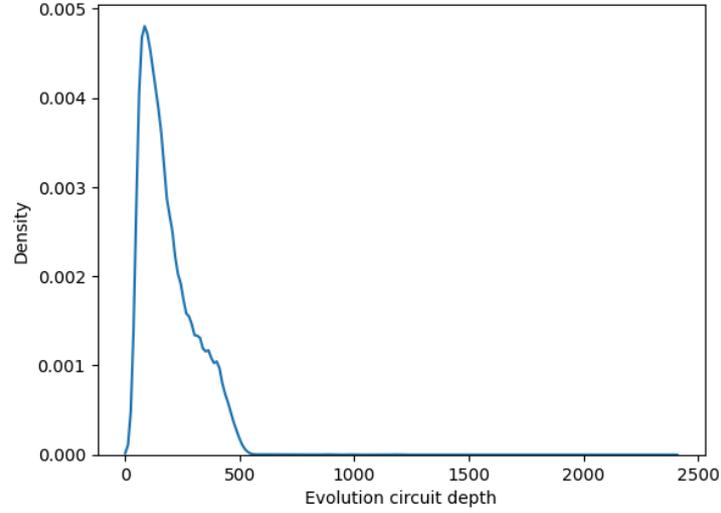

Figure 2: Density plot of baseline evolution circuit depth (i.e. Trotterization of non-reordered Hamiltonians)

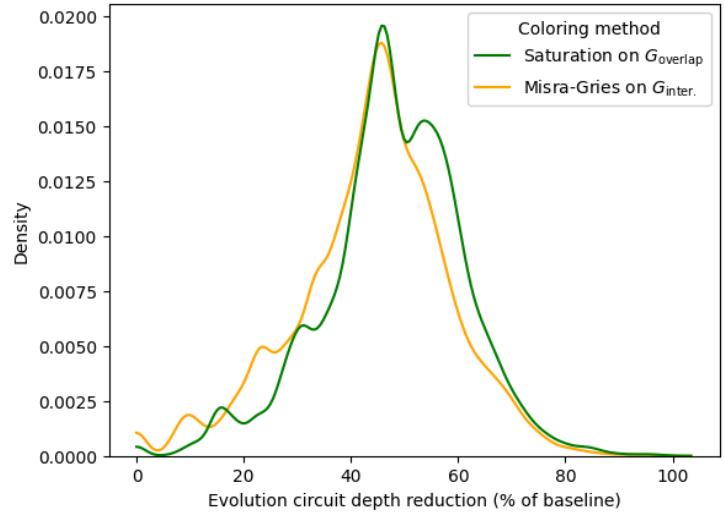

Figure 3: Density plot of the evolution circuit depth reduction after Hamiltonian reordering, relative to baseline

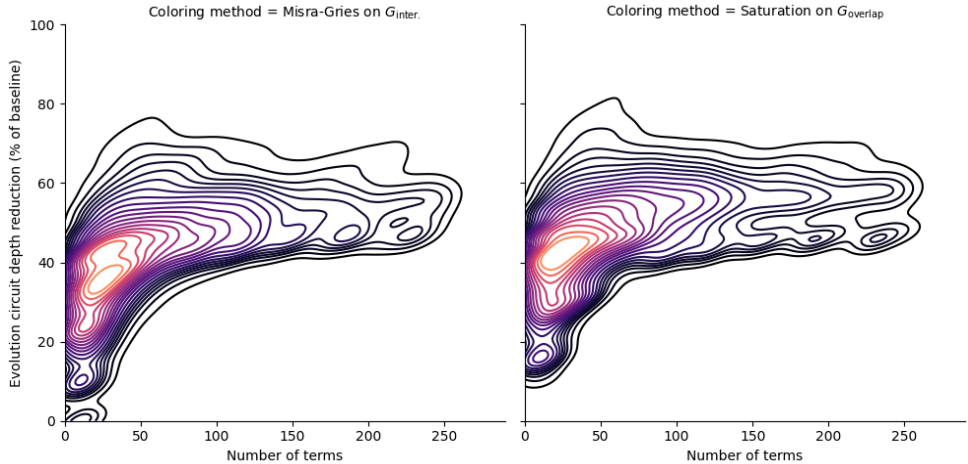

Figure 4: Density plot of the evolution circuit depth reduction against the number of terms in the Hamiltonian

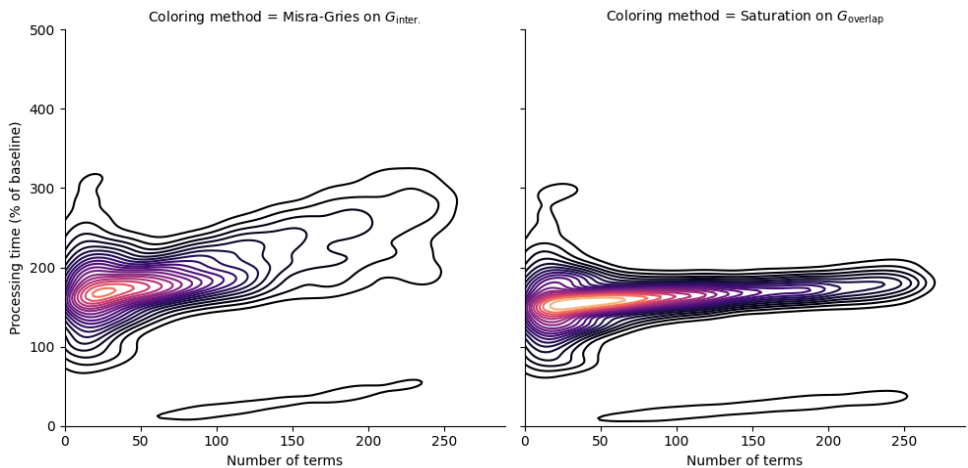

Figure 5: Density plot of the processing time (reordering and Trotterization) relative the the baseline (Trotterization only), against the number of terms in the Hamiltonian

## 4 Application to QAOA

We now demonstrate that shallower evolution circuits as obtained in Section 3 are more conducive to optimization, specifically in the context of QAOA. We compare baseline circuits with circuits derived from reordered Hamiltonians in Section 4.2.

### 4.1 QAOA

In the context of a max-cut problem on a graph $G = (V, E)$, we claimed in Section 3.5 that a highest energy state of the Ising Hamiltonian

$$H_G = -\sum_{(j,k)\in E} Z_j Z_k. \tag{14}$$

describes an maximal cut of $G$. The question is how to find such a state.

QAOA [23–25] is a general variational algorithm designed to approximate a highest energy state of a Hamiltonian. It takes its root in the adiabatic theorem [26], which states that a quantum system remains in a highest energy state if its Hamiltonian is changed "slowly enough". In other words, given two Hamiltonians $B$ and $C$, and $|\psi(0)\rangle$ a highest energy state of $B$, then applying the time-dependent Hamiltonian $H(t) = (1 - t/T)B + (t/T)C$ till a large enough time $T$, will carry the system to the state $|\psi(T)\rangle$ which is a highest energy state of $C$. This process is called *adiabatic evolution* [27, 28].

Simulating a time-dependent Hamiltonian might not be practical or even possible, but this process can be discretized. This gives rise to the Quantum Adiabatic Algorithm (QAA) [27, 29]. Fixing a timestep $\delta t = T/n$ for some large enough $n$, we can approximate $|\psi(T)\rangle$ by $U|\psi(0)\rangle$, where

$$\begin{aligned} U &= \left(e^{-i\beta_{n-1}B} e^{-i\gamma_{n-1}C}\right) \cdots \left(e^{-i\beta_0 B} e^{-i\gamma_0 C}\right) \\ &\approx e^{-i\delta t H(T-\delta t)} \cdots e^{-i\delta t H(0)}, \end{aligned} \tag{15}$$

and where $\beta_k = 1 - k\delta t$ and $\gamma_k = k\delta t$.

QAOA approximates QAA by making the $\beta_k$'s and $\gamma_k$'s variational parameters themselves. In this context, the Hamiltonian $B$ is called the *mixer* and is usually taken to be $B = \sum_{j=1}^{N} X_j$, for which

$$|s\rangle = \left(\frac{|0\rangle + |1\rangle}{\sqrt{2}}\right)^{\otimes N} = \frac{1}{\sqrt{2^N}} \sum_{j=0}^{2^N-1} |j\rangle \tag{16}$$

is a highest energy state, and $C$, the Hamiltonian of interest, is called the *cost operator*. The resulting parametrized operator, also called *QAOA ansatz*, is

$$U(\boldsymbol{\beta}, \boldsymbol{\gamma}) = \left(e^{-i\beta_{n-1}B} e^{-i\gamma_{n-1}C}\right) \cdots \left(e^{-i\beta_0 B} e^{-i\gamma_0 C}\right) \tag{17}$$

which prepares a state $|\boldsymbol{\beta}, \boldsymbol{\gamma}\rangle = U(\boldsymbol{\beta}, \boldsymbol{\gamma})|s\rangle$ carrying an energy $\mathcal{E}(\boldsymbol{\beta}, \boldsymbol{\gamma}) = \langle \boldsymbol{\beta}, \boldsymbol{\gamma} | C | \boldsymbol{\beta}, \boldsymbol{\gamma} \rangle = \langle s | U(\boldsymbol{\beta}, \boldsymbol{\gamma})^\dagger C U(\boldsymbol{\beta}, \boldsymbol{\gamma}) | s \rangle$. This energy is maximized classically over the parameters $\boldsymbol{\beta}$ and $\boldsymbol{\gamma}$ using e.g. the Constrained Optimization By Linear Approximation (COBYLA) algorithm [30–32].

## 4.2 Experiments

We test the benefits of our reordering method on the QAOA algorithm by comparing the energy of the optimal parameters $\mathcal{E}(\boldsymbol{\beta}^*, \boldsymbol{\gamma}^*)$ found during the optimization loop, using the evolution circuits of the baseline and reordered Hamiltonian across various max-cut instances from the HamLib dataset. Here, if $G = (V, E)$ is a graph, then the cost operator is $C = H_G = -\sum_{(j,k) \in E} Z_j Z_k$ as defined in Equation 11.

All experiments are conducted on classical hardware. We use Qiskit version 1.3.1 [33] with the Qiskit Aer 0.15.1 for quantum circuit simulation. The noise model was replicated from the `ibm_kawasaki` device which uses IMB's Eargle R3 chip [34]. The classical minimization loop is implemented using SciPy [35] version 1.15.1 and COBYLA.

The benchmark spans over 3210 max-cut instances, ranging from 7 to 11 edges. The instance sizes are severely constrained by the feasibility of simulating QAOA on a classical device. Running such a large-scale benchmark on actual quantum hardware is currently impractical due to the high demand and the limited number of functional quantum computers.

During each QAOA step, we take interest in the *average energy per edge* (AE/e)

$$\tilde{\mathcal{E}}(\boldsymbol{\beta}, \boldsymbol{\gamma}) = \frac{\mathcal{E}(\boldsymbol{\beta}, \boldsymbol{\gamma})}{|E|} \tag{18}$$

as a key performance indicator. The cumulative maximal AE/e (i.e. the highest AE/e encountered up to a given iteration) of each optimization trial are reported in Figure 6. We observe that reordered Hamiltonians result in better parameters, with a maximal AE/e gain

$$\Delta\tilde{\mathcal{E}} = \tilde{\mathcal{E}}(\boldsymbol{\beta}^*_{\text{reor.}}, \boldsymbol{\gamma}^*_{\text{reor.}}) - \tilde{\mathcal{E}}(\boldsymbol{\beta}^*_{\text{base.}}, \boldsymbol{\gamma}^*_{\text{base.}}) \qquad (19)$$

of $2.33 \times 10^{-3}$ for Hamiltonians reordered using the saturation method, and $1.57 \times 10^{-3}$ for the Misra-Gries method, on average. These gains are reported in Figure 7. The energy gains are modest, most likely due to the small size of the instances considered. We expect $\Delta\tilde{\mathcal{E}}$ to increase for larger instances.

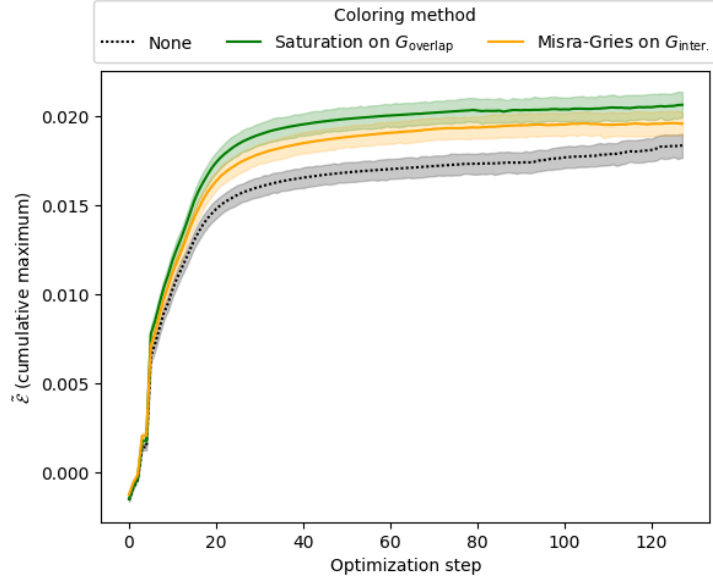

Figure 6: Cumulative maximal AE/e by QAOA step for baseline (black) and reordered (solid green and yellow) Hamiltonians. Lines represent the median, shaded areas represent the 95% confidence interval.

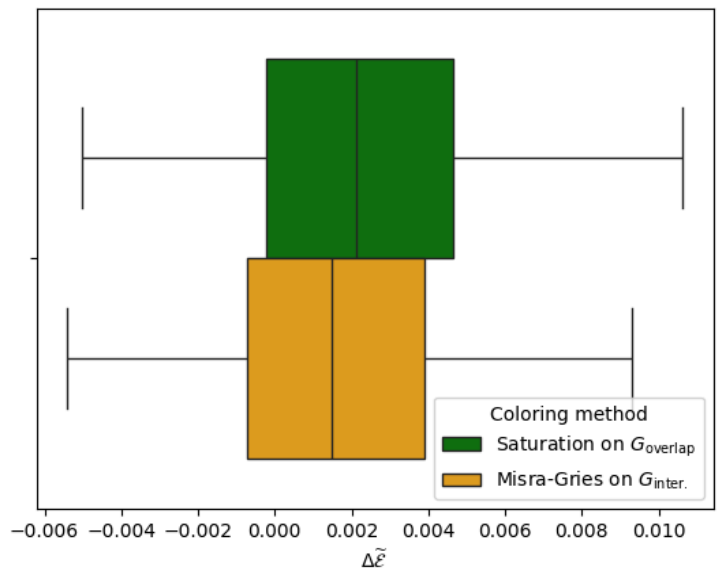

Figure 7: Maximal AE/e's reached during QAOA for reordered Hamiltonians, minus that of the baseline

## 4.3 Assessing the impact of a higher AE/e

A natural next step would be to examine the implication of a higher maximal AE/e on the cut value distribution, i.e. the value of a cut sampled from $|\boldsymbol{\beta}^*_{\text{base.}}, \boldsymbol{\gamma}^*_{\text{base.}}\rangle$ against $|\boldsymbol{\beta}^*_{\text{reor.}}, \boldsymbol{\gamma}^*_{\text{reor.}}\rangle$, where "reor." is the reordering method under consideration. In this section, we suggest several metrics to study this question. For our limited QAOA benchmark, none of these metrics gave a statistically salient distinction between $|\boldsymbol{\beta}^*_{\text{base.}}, \boldsymbol{\gamma}^*_{\text{base.}}\rangle$ and $|\boldsymbol{\beta}^*_{\text{reor.}}, \boldsymbol{\gamma}^*_{\text{reor.}}\rangle$ due to the small sizes of the problem instances that could be processed.

The first and most direct metric is simply the *maximal cut value*
$$\mathcal{M}(\boldsymbol{\beta}, \boldsymbol{\gamma}) = \max\{\text{cut}(c) \mid c \in \text{supp}|\boldsymbol{\beta}, \boldsymbol{\gamma}\rangle\}, \tag{20}$$
Here, we conflate the quantum state $|\boldsymbol{\beta}, \boldsymbol{\gamma}\rangle$ and the induced cut distribution for simplicity. We believe this metric is too coarse in general because if fails to grasp the nature of the cut value distribution induced by $|\boldsymbol{\beta}, \boldsymbol{\gamma}\rangle$. For example, a distribution that exhibits higher cut values consistently may be preferable to a distribution that can find even better cuts but with a low probability, eventhough $\mathcal{M}$ would favor the latter.

If $\mathcal{M}(\boldsymbol{\beta}^*_{\text{base.}}, \boldsymbol{\gamma}^*_{\text{base.}}) = \mathcal{M}(\boldsymbol{\beta}^*_{\text{reor.}}, \boldsymbol{\gamma}^*_{\text{reor.}})$, then the probability of sampling a cut that realizes this bound,
$$\mathbb{P}_{c \sim |\boldsymbol{\beta}, \boldsymbol{\gamma}\rangle}[\text{cut}(c) = \mathcal{M}(\boldsymbol{\beta}, \boldsymbol{\gamma})], \tag{21}$$
can establish a preference between the two distributions. Generalizing this idea, the *average cut value* is naturally given by
$$\mathcal{A}(\boldsymbol{\beta}, \boldsymbol{\gamma}) = \mathbb{E}_{c \sim |\boldsymbol{\beta}, \boldsymbol{\gamma}\rangle}[\text{cut}(c)], \tag{22}$$
It is particularly relevant to distinguish cut distribution whose induced cut value distribution have the same support. However, if $|\boldsymbol{\beta}^*_{\text{base.}}, \boldsymbol{\gamma}^*_{\text{base.}}\rangle$ and $|\boldsymbol{\beta}^*_{\text{reor.}}, \boldsymbol{\gamma}^*_{\text{reor.}}\rangle$ skew heavily towards non-optimal cuts, then $\mathcal{A}$ may fail to point to the better state in a statistically significant manner. Furthermore, if both states reach different but close maximal cut values, i.e. $|\mathcal{M}(\boldsymbol{\beta}^*_{\text{base.}}, \boldsymbol{\gamma}^*_{\text{base.}}) - \mathcal{M}(\boldsymbol{\beta}^*_{\text{reor.}}, \boldsymbol{\gamma}^*_{\text{reor.}})|$ is positive but small, then $\mathcal{A}$ may not clearly reveal this, especially if the range of possible cut values is large.

To restrict our attention to high probability or high cut value samples, we can study the conditional distribution
$$\mathbb{P}_{c \sim |\boldsymbol{\beta}, \boldsymbol{\gamma}\rangle}[\text{cut}(c) \mid c \text{ is Pareto-optimal}], \tag{23}$$
where a sample $c$ is *Pareto-optimal* if there is no other cut $c'$ such that $\mathbb{P}[c'] > \mathbb{P}[c]$ and $\text{cut}(c') > \text{cut}(c)$ simultaneously. In other words, such cuts reach the best tradeoff between likelyhood and cut value. The *average cut value for Pareto-optimal cuts* is then the conditional expectation
$$\tilde{\mathcal{A}}(\boldsymbol{\beta}, \boldsymbol{\gamma}) = \mathbb{E}_{c \sim |\boldsymbol{\beta}, \boldsymbol{\gamma}\rangle}[\text{cut}(c) \mid c \text{ is Pareto-optimal}]. \tag{24}$$

Lastly, we can consider the *hypervolume*, a classical metric from the field of multi-objective optimization, which is the area of
$$\bigcup_c S(\mathbb{P}[c], \lambda \text{cut}(c)), \quad S(x, y) = \{(x', y') \mid 0 \leq x' \leq x, 0 \leq y' \leq y\}. \tag{25}$$

Here, $\lambda > 0$ is a parameter that balances the importance of a higher probability against a higher cut value.

The metrics presented in this section are most likely not computable in practice, but can be empirically estimated given a sufficiently large sample set of cuts.

# 5 Conclusion

This paper revisits the idea of reordering Hamiltonian terms as a preprocessing step, but with the novel goal of minimizing the depth of Trotterized quantum circuits by taking advantage of "gate parallelization" (in the sense of Equation 7). We quantify the efficacy of our method by conducting a large-scale benchmark over a subset of the HamLib dataset.

We then argue that shallower evolution circuits obtained this way are desirable in the context of QAOA. We propose several criteria to assess these benefits, such as increased maximal average energy per edge (AE/e, see Equation 18) and average cut value for Pareto-optimal cuts (Equation 24).

However, due to the high demand and limited number of quantum computers, and the practical limitation of quantum simulation on classical devices, we were not able to conduct a large scale QAOA benchmark to further assess the benefits of our method.